\documentclass[final,3p,times,twocolumn]{elsarticle}

\usepackage{amssymb}
\usepackage{amsmath}
\usepackage{color}

\journal{Chemical Physics}

\begin{document}

\renewcommand{\arraystretch}{1.7}

\begin{frontmatter}



\title{Weighted nodal domain averages of eigenstates for quantum Monte Carlo and beyond}


\author[inst1]{Lubos Mitas}
\ead{lmitas@ncsu.edu}

\author[inst1]{Abdulgani Annaberdiyev}
\ead{aannabe@ncsu.edu}

\affiliation[inst1]{organization={Department of Physics},
            addressline={North Carolina State University}, 
            city={Raleigh},
            postcode={27695-8202}, 
            state={North Carolina},
            country={USA}}



\begin{abstract}
We study the nodal properties of many-body
eigenstates of stationary Schr\"odinger equation that affect the accuracy of real-space quantum Monte Carlo calculations. In particular,
we introduce weighted nodal domain averages that provide a new probe of nodal surfaces beyond the usual expectations. Particular choices for the weight function reveal, for example, that the difference between two arbitrary fermionic eigenvalues is given by the nodal hypersurface integrals normalized by overlaps with the bosonic ground state of the given Hamiltonian. Noninteracting and fully interacting Be atom with corresponding almost exact and approximate wave functions are used to illustrate several aspects of these concepts.
Variational formulations that employ different weights are proposed for prospective improvement of nodes in variational and fixed-node diffusion Monte Carlo calculations.
\end{abstract}

%
%
%
%


\end{frontmatter}



\section{Introduction}

Methods based on stochastic sampling
such as quantum Monte Carlo (QMC) 
proved to be very effective in solving quantum many-body
problems. 
In particular, real-space approaches such as diffusion Monte Carlo (DMC) have been very successful for electronic structure calculations of atomic, molecular, and condensed matter systems \cite{foulkes_quantum_2001, kolorenc_applications_2011, hunt_quantum_2018, benali_toward_2020}. 
The DMC method adopts the so-called fixed-node (FN) approximation in order
to avoid the well-known fermion sign problem
and to provide high accuracy for systems with many degrees of freedom. The fixed-node
 constraint comes from the anti-symmetry of  many-body states that exhibit both positive and negative values separated by a zero hypersurface, i.e., the wave function node.
 The results produced by the fixed-node DMC (FNDMC) from more than three decades are encouraging with many successful research directions for molecules, clusters, transition metal oxide crystals, 2D materials, non-covalently bonded systems, and others \cite{wagner_transition_2007, dubecky_noncovalent_2016, dash_tailoring_2021, huang_bandgaps_2021, wines_first-principles_2020, malone_systematic_2020}. For example, our very recent calculations of 
Si systems show errors between 1.2\% and 3.6\% of correlation energy
and that holds not only for small molecules \cite{wang_binding_2020}
but also for large periodic supercells of Si crystal with hundreds of valence electrons \cite{annaberdiyev_cohesion_2021}. The FNDMC accuracy enabled us not only to produce accurate results but also to offer bounds on both systematic 
and random errors which are smaller than experiments \cite{annaberdiyev_cohesion_2021}. 

There are two key reasons why fixed-node QMC in real space proved to be very successful. The first one is the completeness of the ``basis" represented by the sampling points in the full space of particle coordinates. This avenue has been recently expanded by the sampling of the spin degrees of freedom for treating spin-orbit interactions and beyond in two-component spinor formalism \cite{melton_spin-orbit_2016, melton_quantum_2016}. The possible problems with the ergodicity of sampling in specific cases can 
be overcome by using appropriate generalizations such 
as fixed-phase methods and related \cite{melton_fixed-node_2016, melton_quantum_2017}.
The second key point is that
despite the fixed-node bias, the accuracy of such calculations is remarkably high and in many cases, it is the most accurate many-body wave function
method for large systems. Interestingly, this aspect remains significantly 
less understood and explanations often rely on {\em a posteriori}
justifications, discussions about the used trial wave functions, and a variety of qualitative arguments. {\em A priori} unknown accuracy of nodes for a variety of trial/variational wave functions 
is one of the key obstacles for further progress in this family of QMC approaches.

The goal of this study is to introduce new characteristics of fermion nodes and therefore contribute to
a more systematic understanding of
this unsolved and less investigated problem. 
In particular,
 we show that the nodal behavior of eigenstates encodes information about eigenvalues and it can also provide a possible, if difficult, route for finding the eigenstates themselves.  This understanding is being built through nodal domain averages which we introduced before \cite{hu_many-body_2012} and which we generalize here by introducing a weight function. We explore a few possibilities 
that are offered by this generalization including new identities which reveal that the energy differences between states can be recast as integrals of wave function gradients over nodal surfaces
normalized by overlaps with the bosonic ground state.
These concepts are illustrated on the Be atom that serves as an appropriate testing example since its nodal surfaces for both inaccurate and very accurate ones are well-known and straightforward to construct. We also introduce new variational estimators that take this nodal information into account and therefore open new possibilities for nodal improvements in such calculations. We provide some insights and alternatives in this direction and we speculate about the prospects for Hamiltonians 
with nonlocal terms (such as effective core potentials) within this framework. 



\section{Fixed-node diffusion Monte Carlo}

The fixed-node DMC can be very briefly introduced as follows. We consider a Hamiltonian $H$ for a set of $N$ quantum particles,
for example, a system of interacting electrons and static ions. We approximate its sought after eigenstate 
by a trial or variational  wave function $\Psi_T$. 
Assuming $\Psi_T$ is real, its node is specified by implicit equation $\Psi_T({\bf R})=0$, where particle 
coordinates are denoted as 
${\bf R}=({\bf r}_1, {\bf r}_2, ..., {\bf r}_N)$.

Imposing the trial node on the solution of
the stationary Schr\"odinger equation
can be recast as a delta-function barrier with an infinite strength added to 
the original Hamiltonian \cite{melton_quantum_2016, melton_quantum_2017} as given by 
\begin{equation}
H_{FN} = H+V_{FN}.
\label{eqn:H_eff}
\end{equation}
The DMC algorithm  carries out the projection to the ground state 
\begin{equation}
\Psi_{FN} =\lim_{\tau \to \infty}\exp(-\tau H_{FN})\Psi_T
\label{eqn:psi_proj}
\end{equation}
using the best available $\Psi_T$.
In practice, the projection is carried out by a stochastic process and the fixed-node potential/condition is realized by imposing the
corresponding zero boundary.
The trial functions are typically of the Slater-Jastrow type
\begin{equation}
\Psi_T = e^J \sum_i c_i det_i^{\uparrow}[\varphi_j({\bf r}_k)]
det_i^{\downarrow}[\varphi_l({\bf r}_m)].
\label{eqn:slat_jast}
\end{equation}
Note that the same fixed-node formulation 
applies also to bosonic excited states, namely, the formalism and challenges are essentially the same. Therefore, calculations of bosonic excitations run into the same fundamental difficulties regarding the signs as for fermionic states and  what follows will cover both fermionic and bosonic systems (such as ultracold atoms with effective interactions \cite{li_atomic_2011} and others).

The fixed-node approximation is perhaps the single most important
unsolved problem which hampers the progress in further improvement of
 accuracy and efficiency of many QMC calculations, and, in fact, even beyond, since the challenges of signs (as well as complex amplitudes) appear in simulations of many-body quantum system  in general.
 Experience from electronic structure structure calculations provide many illustrations that show how FNDMC results are determined by the quality of
the trial wave function nodes. Over the years, a significant amount of effort 
has been devoted to trial function improvements \cite{benali_toward_2020, giner_using_2013, lopez_rios_inhomogeneous_2006, bajdich_pfaffian_2006, bajdich_pfaffian_2008, townsend_starting-point-independent_2020, luchow_direct_2007, toulouse_optimization_2007}. 
Despite these investments,
 our knowledge about the nodes remains rather limited since it is closely related
 to fundamental mathematical problems (indeed, nodal topologies and nodal domain counts of eigenstates belong to the celebrated problems proposed by David Hilbert at the beginning of 20-th century \cite{hilbert_mathematical_1902}).
 
One of the key difficulties is that the fixed-node bias is very small
on the scale of total energies of electron-ion systems with Coulomb interactions. It is well-known that the correlation energy,
defined as the difference between the exact and the Hartree-Fock (HF) energies, is typically a few percent of the total energy. As found out by many a posteriori estimations from QMC data, the fixed-node error is typically a few percent of the correlation energy, so that we end up with $\approx$ {\em 0.1\%} of the total energy. 
Of course,
even this level of bias can determine whether one is able to study important classes of quantum phenomena or not. 
The true energetic impact is very difficult to discern, in particular in variational Monte Carlo (VMC) methods that are routinely used to optimize the trial functions since the overall level of statistical noise is driven by the largest energy scale present, such as the potential energy. In addition, such optimizations are, in general, still very tedious and in many cases just too costly.

In the previous paper \cite{hu_many-body_2012}, we have tried to look at the nodal biases from a new angle.
In order to shed some light on such cases,
we suggested so-called nodal domain 
averages  that actually enabled us to reveal new aspects of node properties in a quantitative
manner. Here we make another step in this direction with new insights into the node characteristics
and their connections to eigenvalues, eigenstates and related expectations of interest. 

\section{Nodal domain averages}

Let us recall our recent results \cite{hu_many-body_2012} where we have shown that
for any exact fermionic (or excited bosonic)
eigenstate one can express the total energy
 as a sum of kinetic and potential components, that we dubbed nodal domain
averages (nda, in short). They are defined as follows
\begin{equation}
E_{kin}^{nda}=\int_{\partial\Omega}|\nabla_{\bf R}\Psi({\bf R})|\text{d}{\bf S}/
\int |\Psi({\bf R})| \text{d}{\bf R},
\label{eqn:nda_kin_def}
\end{equation}
and
\begin{equation}
E_{pot}^{nda}=\int V({\bf R})|\Psi({\bf R})|  \text{d}{\bf R}/
\int |\Psi({\bf R})| \text{d}{\bf R},
\label{eqn:nda_pot_def}
\end{equation}
so that
\begin{equation}
E=E_{kin}^{nda}+E_{pot}^{nda}
\label{eqn:nda_total}
\end{equation}
while we assumed $H\Psi=E\Psi$.
This deserves some comments as it applies to {\em any eigenstate}
including excitations, both fermionic and bosonic (the bosonic ground state
is a trivial exception since it is generically nodeless).
The most interesting property of these expressions is that they probe eigenstates from a different  perspective than usual ``two-sided" $\langle \Psi |H| \Psi \rangle$ expectations. 
The aim is to measure 
the impact of the node on the total energy more directly from the wave function amplitude. Note that in usual  expectations the node contributions are very indirect since at the node both
the square of the eigenstate and its Laplacian vanish, hence, the weak ``signal".
This is just another indication that the fixed-node error 
in reality can turn out to be small and in some cases surprisingly small even for very crude nodal guesses. As we also mentioned originally,
the nda
expression is not necessarily variational since it is not quadratic in the
wave function as it is for the usual expectation value of the Hamiltonian. It is only ``one-sided"
and as such both potential, kinetic,
and total energy components deviate to the first order in wave function error.
We consider this point and elaborate on it further in more detail
later.

We showed that nda quantities enabled us to obtain interesting insights into the nodal properties as well as to find equivalence between the nodes
that were completely unexpected. Let
us mention a few examples: 

i) The nda components differ
for degenerate eigenstates. For example,
building upon previous work \cite{hu_many-body_2012}, we were able to analytically derive that for the noninteracting Be atom ground and excited state the components show clearly different values (in previous work the values for the state $^1S(1s^22p^2)$ were estimated numerically).
These values are shown in Table \ref{tab:be_noni}.


\begin{table}[!htbp]
\centering
\caption{
Energy components (a.u.) of noninteracting Be atom ground and excited states: ordinary expectations and nda values.
}
\label{tab:be_noni}
\begin{tabular}{lcccccc}
\hline
\hline

State  & E$_{tot}$ & E$_{kin}$ & E$_{pot}$ & E$_{kin}^{nda}$ & E$_{pot}^{nda}$ \\
\hline
$^1S(1s^22s^2)$ & -20  & 20 & -40 & $\frac{320}{221}$ & $-\frac{4740}{221}$ \\
$^1S(1s^22p^2)$ & -20  & 20 & -40 & $\frac{8}{5}$     & $-\frac{108}{5}$    \\

\hline
\hline
\end{tabular}
\end{table}

Therefore
one can see a clear signature of the different nodes on nda components while the usual expectations are all identical. It is interesting that for an interacting Be atom, the mix of these two states provides a very accurate interacting node 
using trial function based on HF calculation with added double excitation
\begin{equation}
\Psi_{\rm 2-conf} = \Psi(1s^22s^2)-c_0\Psi(1s^22p^2)
\label{eqn:be_exact}
\end{equation}
as demonstrated previously
\cite{umrigar_diffusion_1993}. This trial function 
will be further employed later as the  the determinantal part of trial function for 
interacting Be atom.

ii) We found that fermionic 
and bosonic nodes could be equivalent modulo some coordinate transformation. 
Consider three noninteracting states $^3P(2p^2)$, $^1D(2p^2)$, and $^1S(2p^2)$ in attractive spherical potential. It turns out 
that the nda kinetic and potential components as well as the usual kinetic and 
potential energy components are degenerate, showing that the nodes are equivalent.
The fact that such different spatial and spin symmetries lead to
equivalent  nodal properties came out as a surprise. In addition, for example, the node 
of the $d_{2z^2-x^2-y^2}$ state shows ``accidental 
degeneracy" since its geometry
is demonstrably 
different from the rest of the $d$-subshell (say, from $d_{xz}$). Similar observations 
about nodal properties of 
spherical harmonics were noticed also 
elsewhere \cite{sogge_2011}.

\section{Weighted nodal domain averages}

After introducing the nodal domain averages
\cite{hu_many-body_2012}, we later found that the nda expressions correspond to a special case of identity which was derived idependently by
Sogge and Zelditch
\cite{sogge_2011}. We will present here the identity in a form that includes potentials which is a straightforward generalization 
(the original paper \cite{sogge_2011} assumes
Laplacian eigenstates, i.e., with no potentials). Since many properties will apply equally to both fermionic {\em and} bosonic excited states that necessarily exhibit nodes as well (except the nodeless bosonic ground state), we will consider these two classes of symmetries together unless specified otherwise. 

The stationary Schr\"odinger equation 
for an eigenstate $\Psi$ with an eigenvalue $E$ is given by
\begin{equation}
T_{kin}\Psi({\bf R}) +V({\bf R})\Psi({\bf R})=E\Psi({\bf R})
\label{eqn:stat_schr}
\end{equation}
where $T_{kin}=
-(1/2)\sum_i\nabla_i^2 $. 
For simplicity, we assume free boundary conditions for bound normalizable states or a finite volume (crystal) cell with periodicity. 
An exact fermionic (or excited bosonic)
real eigenstate $\Psi$
exhibits the nodal domains
\begin{equation}
\Omega^+_{\Psi}=\{{\bf R}; \Psi({\bf R})>0\}, \; \Omega^-_{\Psi}=\{{\bf R}; \Psi({\bf R})<0\}
\label{eqn:nodal_domains}
\end{equation}
and the corresponding node $\partial\Omega_{\Psi}$ which is implicitly given by $\Psi({\bf R})=0$.
We proceed by multiplying the Schr\"odinger equation by 
a real function $\Phi({\bf R})$
that we call simply a weight. 
We will consider weights that are in general non-negative although other options are, in general, possible.
We integrate by parts twice over the $\Omega^+_{\Psi}$ domain {\em } (Green's theorem) so that we obtain
$$
-\frac{1}{2}\int_{\partial\Omega_{\Psi}}\Phi({\bf R})\nabla_{\bf R}\Psi({\bf R})\cdot \text{d}{\bf S}_n+
\int_{\Omega^+_{\Psi}} \Phi({\bf R})V({\bf R})\Psi({\bf R}) \, \text{d}{\bf R}+
$$
\begin{equation}
+\int_{\Omega^+_{\Psi}} \Psi ({\bf R})T_{kin}\Phi({\bf R})\text{d}{\bf R} =
E\int_{\Omega^+_{\Psi}} \Phi({\bf R})\Psi({\bf R}) \text{d}{\bf R}
\label{eqn:int_twice}
\end{equation}
where ${\bf S}_n=-\nabla\Psi/|\nabla\Psi|$ is the normal 
vector to the node. 
We assume that all the needed integrals exist and the analytical properties of all involved functions are such that one can follow the proof steps \cite{sogge_2011}
in a straightforward manner.
Note that if there is more than one
domain of the same sign, we can repeat this for each such domain. 
Subsequently, we carry out the same   for  $\Omega^-$ domain(s) and by subtracting the two equations we obtain
\begin{multline}
\int_{\partial\Omega}\Phi({\bf R})|\nabla_{\bf R}\Psi({\bf R})| \text{d}{\bf S}
+\int V({\bf R})\Phi({\bf R})|\Psi({\bf R})| \text{d}{\bf R}+ \\
+\int |\Psi ({\bf R})|T_{kin} \Phi({\bf R})\text{d}{\bf R}
=E\int \Phi({\bf R})|\Psi({\bf R})| \text{d}{\bf R}.
\label{eqn:phi_sub}
\end{multline}
We assume that the function $\Phi({\bf R})$ is such that the functions $\Phi$ and $|\Psi|$
are not orthogonal. In particular, $\Phi$ should not be fermionic (antisymmetric) since then the overlap integral trivially vanishes. Also, when $\Phi = |\Psi|$ the integral over $\partial\Omega$ becomes zero  and we recover 
the usual expectation for the total energy. On the other hand, for otherwise arbitrary eigenstate $\Psi$ the 
obtained equation shows that the total energy
is given as a sum of kinetic and potential components weighted by $\Phi$ plus the Laplacian contribution of $\Phi$. That leads us to the definition of the following {\em $\Phi-$weighted nodal domain averages} 
\begin{equation}
E_{kin}^{nda}
=\int_{\partial\Omega}\Phi({\bf R})
|\nabla_{\bf R}\Psi({\bf R})|\text{d}{\bf S}/
\int \Phi({\bf R}) |\Psi({\bf R})| \text{d}{\bf R},
\label{eqn:wnda_kin}
\end{equation}

\begin{equation}
E_{pot}^{nda}=\int V({\bf R})
\Phi({\bf R})
|\Psi({\bf R})|  \text{d}{\bf R}/
\int \Phi ({\bf R})|\Psi({\bf R})| 
\text{d}{\bf R},
\label{eqn:wnda_pot}
\end{equation}

\begin{equation}
E_{kin,\Phi}^{nda}=\int |\Psi({\bf R})|
T_{kin} \Phi({\bf R})
  \text{d}{\bf R}/
\int \Phi({\bf R})|\Psi({\bf R})| 
\text{d}{\bf R}
\label{eqn:wnda_kin_phi}
\end{equation}
and emphasizing the role of the weight $\Phi$ we write
\begin{equation}
E^{nda}(\Phi)=E_{kin}^{nda}(\Phi)
+E_{pot}^{nda}(\Phi)+
E_{kin,\Phi}^{nda}.
\label{eqn:wnda_tot_phi}
\end{equation}

In our previous work \cite{hu_many-body_2012} on nda averages  (Eqs.(4-6)) 
the weight function was simply a constant $\Phi=1$. In what follows we will use the same simple notation assuming that
the weight function will be obvious from the context, ie,
$E^{nda}(\Phi)\to E^{nda}$, etc.

The freedom of choice of the weight function opens new possibilities since one can vary $E_{kin}^{nda}, E_{pot}^{nda}$ and possibly even eliminate some term(s)
by using an appropriate weight function. Let us consider, for example, Coulomb interactions in an ion-electron system so that $V$ has both positive and negative values. There is an option to choose 
the weight in such a manner that the kinetic 
energy term vanishes. In such  case the exact energy is given solely by the integral over the node, 
potential energy of the weight function 
$\Phi$, and overlap $\langle \Phi||\Psi|\rangle$.  Clearly, the components will vary from state to state and corresponding weights would have to vary accordingly.
However, there is a possibility to fix the weight for all states so that the only remaining integrals over the whole configuration space are overlaps, as we will explain in what
follows.

\subsection{One-particle product as $\Phi$} 
Let us 
consider the weight given by a product of one-particle functions
\begin{equation}
\Phi({\bf R})=\prod_{i=1}^{N}\eta({\bf r_i})
\label{eqn:w_spo}
\end{equation}
where the orbital $\eta({\bf r})$ is given by a
single-particle Schr\"odinger equation  with appropriate effective 
potential $V_0$ 
\begin{equation}
[-(1/2)\nabla^2_{\bf r}+V_0({\bf r})-e_0]\eta({\bf r})=0.
\label{eqn:schr_spo}
\end{equation}
We assume that this equation is analytically or numerically solvable so that we can write $e_{\Phi}=Ne_0$, $V_{\Phi}=\sum_iV_0({\bf r}_i)$ and it follows
that the eigenvalue is given by 
\begin{equation}
E=\frac{\int_{\partial\Omega}\Phi ({\bf R})
|\nabla_{\bf R}\Psi({\bf R})|\text{d}{\bf S}}
{
\int \Phi({\bf R}) |\Psi({\bf R})| \text{d}{\bf R}
}
+e_{\Phi}+\langle V-V_{\Phi}\rangle_{\Phi|\Psi|}
\label{eqn:wnda_no_kin}
\end{equation}
which simplifies the nda expression by eliminating the kinetic energy of the weight function $\Phi$. 
Furthermore, this form 
looks reasonably practical since it is rather straightforward to produce one-particle functions for $\Phi$
that will have sufficiently large overlap with $|\Psi|$. 
It is also tempting to think about the possibilities that would lead to vanishing last term by an appropriate tuning of $V_0$.

\subsection{Bosonic ground state as the weight function}
An intriguing choice is to set $\Phi$ equal to
the bosonic ground state of the original Hamiltonian
\begin{equation}
\Phi({\bf R})=\Phi_{0B}({\bf R})
\label{eqn:boson_gs}
\end{equation}
Then the energy of a given state, fermionic or excited bosonic,
 is given solely by the integral over the nodal surface 
and the overlap integral that provides the corresponding normalization
\begin{equation}
E=\frac{\int_{\partial\Omega}\Phi_{0B}({\bf R})
|\nabla_{\bf R}\Psi({\bf R})|\text{d}{\bf S}}
{
\int \Phi_{0B} ({\bf R}) |\Psi({\bf R})| \text{d}{\bf R}
}
+E_{0B}
\label{eqn:en_bosonic}
\end{equation}
This is a simplified version of the above Eqn. \ref{eqn:wnda_no_kin}.
Clearly, the expression shows what is the minimal knowledge necessary for recovering the eigenvalue and the eigenstate, fermionic or excited bosonic:  node, gradient at the node, and overlap. 
This is straightforward to understand since the node and the gradient provide initial conditions for solving the second-order differential (Schr\"odinger) equation while the overlap fixes the normalization/eigenvalue of the corresponding bound state. 
In addition, we recognize a special type of holography since such simple expression does not hold for any other level set of the eigenstate, only for the node. 
Trivially, for excitation energies  $E_{gap}=E_{ex}-E_{gr}$ we get
\begin{equation}
E_{gap}=
\frac{\int_{\partial\Omega_{ex}}\Phi_{0B}
|\nabla_{\bf R}\Psi_{ex}|\text{d}{\bf S}}
{
\int \Phi_{0B} |\Psi_{ex}| \text{d}{\bf R}
} - 
\frac{\int_{\partial\Omega_{gr}}\Phi_{0B}
|\nabla_{\bf R}\Psi_{gr}|\text{d}{\bf S}}
{
\int \Phi_{0B}|\Psi_{gr}| \text{d}{\bf R}
}.
\label{eqn:gap}
\end{equation}
Note that the potentials and interactions
do not appear explicitly, their impact is implicitly encoded into the quantities that enter the  equation. Let us also comment on the fact that for obtaining 
information about fermionic states we would need also the bosonic ground state.  Although the bosonic eigenstate can be rather challenging to find in general, this plays into the strength of the QMC methods since $\Phi_{0B}$ is nodeless. Therefore it is possible to obtain samples of the exact solution and its energy without the fundamental problem of fermion signs in straightforward DMC calculations. In this sense, the expressions with the exact bosonic state
above are particularly interesting in the context of QMC methods. 

 The simplicity of the expressions
 is rather suggestive in several aspects 
and it is tempting to
reinterpret it as follows. The bosonic ground state carries an imprint of the external potential and interactions, and we can consider it to be its nonnegative envelope  function while its eigenvalue provides an overall reference level
for the spectrum.
This suggests that we could write the many-body fermionic states as a product of the bosonic ground state envelope and the antisymmetry component that would be given as  an expansion in a particular basis (the simplest such basis that comes to mind are multivariate
anti-symmetric polynomials).  This form is exact for noninteracting harmonic fermions (for both symmetric and antisymmetric states) and also
an excellent approximation for the same system with interactions.

The Slater-Jastrow wave functions, which has been used in many studies of interacting quantum problems and became very prominent in QMC, qualitatively correspond to this model since it is
trivial to rewrite it as
\begin{equation}
\Psi =e^J\left[\prod_i\rho({\bf r}_i)\right] det[\{\varphi_j({\bf r}_i)/\rho({\bf r}_i)\}]
\label{eqn:slat_jast_bos}
\end{equation}
where $\rho({\bf r})$ is the one-particle 
density so that the first two factors approximate the bosonic envelope.
 An important additional complication for Coulomb potentials is that each subshell
requires its own density factor, implying that
the functions in the determinant are not polynomials but products of polynomials and exponentials with varying exponents. Close to the nucleus the factor is $e^{-Zr/n}$ where $n$ is the principal quantum number while the tail exponent is determined by the corresponding subshell
ionization potential.
Note that the traditional expansions appear both slowly converging and also difficult to optimize
due to the points mentioned earlier. 
The expressions above  expose the nodal properties more explicitly and therefore could provide new ways how to improve the nodal approximations.

\subsection{Toy model example} 
For a single particle in Coulomb potential  $V=-Z/r$ we consider $p-$state with the node $\Psi_{2p}=\exp(-Zr/2)z$
and the exact 
ground state as a bosonic eigenstate or weight $\Phi=\exp(-Zr)$. Considering 
Eq.\ref{eqn:en_bosonic} we find
\begin{equation}
E^{nda}=E_{2p}= E_{kin}^{nda} -E_{1s}=
3Z^2/8-Z^2/2=-Z^2/8
\label{eqn:toy_model}
\end{equation}
as expected.


\subsection{Non-variational vs variational formulation} 
Let us expand the nda formulation to non-exact, variational and QMC framework.
First, it is straightforward to show that,
in general, the 
weighted nodal domain expressions
are not variational per se. 
We assume 
$\Psi_0$ to be the fermionic ground state and we expand 
 the variational state $\Psi_T$ in exact eigenstates 
$
\Psi_T=c_0\Psi_0+\sum_{i>0}c_i\Psi_i.
$
Integrating over one domain and assuming the corresponding normalizations, we find
\begin{equation}
\int_{\Omega_{\Psi^+_T}}\Phi({\bf R})
H\Psi_T({\bf R})d{\bf R}= E_0+\sum_{i>0}c_id_i\Delta E_i
\label{eqn:en_expand}
\end{equation}
where
\begin{equation}
d_i=
\int_{\Omega_{\Psi^+_T}}\Phi({\bf R})
\Psi_i({\bf R})d{\bf R};\; \Delta E_i= E_i-E_0.
\label{eqn:d_overlap}
\end{equation}
The signs of the coefficients $\{c_id_i\}$ vary, in general, so that the resulting energy can be below $E_0$.

On the other hand, it is not too difficult to construct $\Phi$  and $\Psi_T$ so that the nda expression is variational. 
We adapt the nda derivation  by considering the modified Schr\"odinger equation that involves approximate $\Psi_T$
 as follows
\begin{equation}
H\Psi_T({\bf R})= E_L({\bf R}) \Psi_T({\bf R})
\label{eqn:local_fn}
\end{equation}
where the local energy $E_L({\bf R})=
[H\Psi_T({\bf R})]/\Psi_T({\bf R})$ is a function, not an eigenvalue. On the left side we repeat the same steps that we used in obtaining Eq.\ref{eqn:wnda_kin} - \ref{eqn:wnda_kin_phi} while replacing $\Psi$ with $\Psi_T$
\begin{equation}
E_{kin}^{nda}
=\int_{\partial\Omega_{\Psi_T}}
\negthinspace\negthinspace\Phi
|\nabla_{\bf R}\Psi_T|\text{d}{\bf S}\,/\negthinspace
\int \Phi |\Psi_T| \text{d}{\bf R},
\label{eqn:nda_kin_T}
\end{equation}
etc.
The right side is then adapted by dividing and multiplying by the overlap as given by
\begin{equation}
\int \Phi E_L |\Psi_T| d{\bf R} =\langle E_L\rangle_{\Phi|\Psi_T|}
\int \Phi|\Psi_T| d{\bf R}.
\label{eqn:write_ovlp}
\end{equation}
 This can be further reformulated also for DMC estimator using the fixed-node solution  $\Psi= \Psi_{FN}$.
The fixed-node DMC estimator requires
a minor generalization since 
the gradients of $\Psi_{FN}$
are not continuous 
across the node. Therefore, the corresponding expression Eq.\ref{eqn:nda_kin_T} is modified
as follows 
\begin{equation}
E_{kin}^{nda} =
\frac{\int_{\partial\Omega}\Phi({\bf R})
[|\nabla_{\bf R}\Psi({\bf R})|_+
+|\nabla_{\bf R}\Psi({\bf R})|_-]
\text{d}{\bf S}}
{
2\int \Phi ({\bf R}) |\Psi({\bf R})| \text{d}{\bf R}
}.
\label{eqn:dmc_nda}
\end{equation}
The subscript signs indicate whether the gradients are evaluated from the positive or from the negative side. 
This also shows that the key improvement from variational to FNDMC is coming from the shape of the wave function that also influences its gradient at the node. 

Note that the corresponding computational implementation in QMC methods is straightforward since one simply calculates another set of averages. If we now consider trial state optimized by variational QMC methods we can calculate the nda components and the energy directly while choosing, say, the most simple $\Phi=1$. We can expect that the corresponding nda value will be typically 
{\em higher} than the exact energy. The qualitative reasoning that applies rather broadly can be understood as follows. The trial state is optimized with regard to the VMC energy that is given by the integral of the local energy weighted with $\Psi_T^2$. 
In the nda expressions with $\Phi=1$  this weight 
changes to $|\Psi_T|$ and that reduces contributions from regions where the local energy is close to the exact eigenvalue or possibly lower in some regions. On the other hand, this increases the weight in 
regions where (on average) the local energy is higher.
In order to illustrate this, we present calculations of the interacting Be atom as a testing example. The Table \ref{tab:be_interact} shows  single-reference Hartree-Fock,
two-reference (no Jastrow), two-reference Slater-Jastrow VMC, and the corresponding fixed-node DMC 
expectation values. Clearly, the nda total energy estimator does not look very useful, at least not in its raw form. It is significantly higher than the commonly used expectations and this has to be the case since ordinary expectations converge quadratically while the nda expectations show a much larger linear order of deviations. Considering however, that we are focused on amplifying the nodal error signal, this is potentially promising 
since here it has a chance to be significantly stronger than in commonly used variational calculations (we recall that the resulting fixed-node errors are typically $\approx$ 0.1\% of the total).
In addition, we can identify the direct contribution from the nodal surface and therefore could build a different variational scheme that would target the
nodal shape of the trial function. The Table \ref{tab:be_interact} shows that even a rather crude trial function can exhibit exact or nearly exact nodes. In order to pave the way for such direction,  we now
illustrate that it is possible to adapt the variational formulation that 
includes the nda contributions.

%
%
%
%
%

\begin{table*}[!htbp]
\centering
\caption{
Energy (a.u.) components of interacting Be atom ground state: nda components and total energy with $\Phi=1$ and ordinary expectation values. For the first two rows, WF denotes HF and 2-conf trial functions, both without Jastrow factors. 
VMC and DMC energies are obtained with 2-conf Slater-Jastrow trial wave function.
}
\label{tab:be_interact}
\smallskip
\begin{tabular}{lllllll}
\hline
\hline

WF  & E$_{kin}^{nda}$ & E$_{pot}^{nda}$ & E$^{nda}$ & $\langle H \rangle$ \\
\hline
HF      & 1.126(6) & -15.248(6) & -14.1214(4) & -14.5731(2) \\
2-conf  & 1.141(5) & -15.281(6) & -14.1399(5) & -14.6129(3) \\
VMC     & 1.191(9) & -15.820(9) & -14.6290(2) & -14.6627(1) \\
DMC     & 1.221(9) & -15.862(9) & -14.6409(4) & -14.6670(1) \\
Exact \cite{stanke_five_2009}
        &          &            &             & -14.667356486 \\

\hline
\hline
\end{tabular}
\end{table*}


\subsection{Weighted VMC and nda variational formulation}
Let us consider
\begin{equation}
\Phi= |\Psi_T| + \alpha 
\label{eqn:phi_alpha}
\end{equation}
where $\alpha$ is not too large. We expect that this will be universally variational since for small $\alpha$ we get $c_i \approx d_i$, namely, an upper bound regime. 
After multiplying with $\Phi$, integrating and rearranging, the
Eq. \ref{eqn:local_fn}  can be expressed as
a weighted variational combination
\begin{equation}
E_{VMC,nda} =\langle H \rangle_{\Psi_T^2}
+\frac{\alpha\omega}{1+\alpha \omega}
[E^{nda}(\Phi=1)-\langle H\rangle_{\Psi_T^2}]
\label{eqn:vmc_nda}
\end{equation}
where 
\begin{equation}
\omega=\int |\Psi_T| d{\bf R}\; /
\negthinspace\int |\Psi_T|^2 d{\bf R}
\label{eqn:omega_ratio}
\end{equation}
and we emphasize that the involved $E^{nda}$ is evaluated with $\Psi=\Psi_T$ and $\Phi=1$.
As we claimed and illustrated above 
in the Table \ref{tab:be_interact}, the difference in the square bracket is expected to be positive. 
Note that this difference is easy to evaluate as a byproduct of VMC calculation and therefore this can be used to cover for the possibility that this becomes negative. In the case for a given state this difference would come out as negative, 
one can choose 
$\alpha$ negative as well while small enough to guarantee
$1+\omega\alpha>0$. Therefore, for small $\alpha$ the linear combination above can be driven into a variational regime essentially automatically for {\em any} choice of $\Psi_T$. We also emphasize that
the presented choices including  Eq. \ref{eqn:phi_alpha} have been motivated mainly by analytical convenience and many
more choices are at our disposal.


\subsection{Nodal domain averages and nonlocal Hamiltonians}
Consider a Hamiltonian that, in addition to local potentials and interactions includes also a nonlocal term
\begin{equation}
H=T+V+W;  \quad
W=\int w({\bf R},{\bf R}') d{\bf R}'
\label{eqn:nl_def}
\end{equation}
as is the case for effective core potentials \cite{dolg_relativistic_2012} or other generalizations.
From the point of view of nodes, there is an important difference with local only Hamiltonians. At the node 
the Laplacian of the 
wave function should vanish, however,
this is not true in general for 
Hamiltonians with off-diagonal terms. In fact, it is the 
sum of the kinetic and nonlocal term(s)
\begin{equation}
T\Psi+\int w({\bf R},{\bf R}')\Psi({\bf R}') d{\bf R}' =(E-V)\Psi
\label{eqn:nl_vanish}
\end{equation}
that vanishes at the nodal
boundary. The effect of this is opening a new ``variational freedom
for the Laplacian" and therefore one can expect a more significant impact (more mixing) of excitations in the
corresponding many-body eigenstate. 
The nodal domain averages will get
 an additional 
new contribution that can be written as
$$
E_W^{nda}=\langle\Phi|W|\Psi\rangle/
\int \Phi({\bf R})|\Psi({\bf R})|
\text{d}{\bf R}=
$$
\begin{equation}
=\int\Phi({\bf R}) |\Psi({\bf R})|
\tilde W ({\bf R}) d{\bf R}/
\int \Phi({\bf R})|\Psi({\bf R})| 
\text{d}{\bf R}
\label{eqn:nl_nda}
\end{equation}
where 
\begin{equation}
\tilde W ({\bf R})= \int w({\bf R},{\bf R}')\Psi({\bf R}')/|\Psi({\bf R})|d{\bf R}'.
\label{eqn:nl_loc}
\end{equation}
The analysis therefore becomes more involved and 
considering the scope of this work, we do not pursue this aspect any further.  Clearly, this requires more analysis which we leave for future work.

\section{Conclusions}
In this work we provide a theoretical analysis
of connections between the eigenvalues and nodes of many-body eigenstates of stationary Schr\"odinger equation.
We introduce {\em weighted nodal domain averages} that advance our previous work where the weight function was assumed to be constant. The derived formulas provide relations for exact eigenvalues of any real fermionic or excited bosonic eigenstate based only on the nodal surface integral of the eigenstate gradient, the bosonic ground state
energy, and the overlap of its absolute value with the bosonic ground state. The energy differences involve only 
the difference of such normalized integrals over the nodal surface. The expression does not involve the potentials, i.e., potential and interactions  enter only implicitly through their imprint on the mentioned quantities. For the energy difference between the states even the bosonic eigenvalue drops out, namely, one gets the  excited spectrum solely
from the nodal subspace integrals and
overlaps. 

We explore the possibilities how these relations could provide new opportunities 
for variational improvements of nodes in combination with variational Monte Carlo
approaches. We illustrate some of these aspects on Be atom, both with exact analytical results for its noninteracting version as well as with an accurate VMC trial function that
in fixed-node DMC provides nearly exact 
total energy.

Very recently, we explored generalization from the fixed-node  to fixed-phase formalism that applies to complex wave functions and where, seemingly, 
the nodes do not play a prominent role \cite{melton_quantum_2017}.
However, the corresponding formalism relies on an approximate phase that is fully determined by the real and imaginary parts of the wave function. Since these components indeed have nodes, the issue of node accuracy
is very much present in this generalization as well.
Finally, we mention complications that arise from 
nonlocal terms in the Hamiltonian and how the corresponding expression changes with further analysis left for future work.


We believe the work opens new possibilities for 
studies of accuracy of nodal surfaces 
for both fermionic
and bosonic systems and for improving the 
nodal hypersurface approximations.
The practical implementations require
new developments on the software side and corresponding thorough testing that will be reported in future.


\section{Declaration of Competing Interest}
The authors declare that they have no known competing financial interests or personal relationships that could have appeared to influence the work reported in this paper.

\section{Acknowledgments}

This research was supported by the U.S. Department of Energy (DOE), Office of Science, Basic Energy Sciences (BES) under Award DE-SC0012314. 

%
%
\bibliographystyle{elsarticle-num} 
\bibliography{main}

\begin{thebibliography}{10}
\expandafter\ifx\csname url\endcsname\relax
  \def\url#1{\texttt{#1}}\fi
\expandafter\ifx\csname urlprefix\endcsname\relax\def\urlprefix{URL }\fi
\expandafter\ifx\csname href\endcsname\relax
  \def\href#1#2{#2} \def\path#1{#1}\fi

\bibitem{foulkes_quantum_2001}
W.~M.~C. Foulkes, L.~Mitas, R.~J. Needs, G.~Rajagopal, Quantum {Monte} {Carlo}
  simulations of solids, Rev. Mod. Phys. 73~(1) (2001) 33--83.
\newblock \href {https://doi.org/10.1103/RevModPhys.73.33}
  {\path{doi:10.1103/RevModPhys.73.33}}.

\bibitem{kolorenc_applications_2011}
J.~Kolorenč, L.~Mitas, Applications of quantum {Monte} {Carlo} methods in
  condensed systems, Rep. Prog. Phys. 74~(2) (2011) 026502.
\newblock \href {https://doi.org/10.1088/0034-4885/74/2/026502}
  {\path{doi:10.1088/0034-4885/74/2/026502}}.

\bibitem{hunt_quantum_2018}
R.~J. Hunt, M.~Szyniszewski, G.~I. Prayogo, R.~Maezono, N.~D. Drummond, Quantum
  {Monte} {Carlo} calculations of energy gaps from first principles, Phys. Rev.
  B 98~(7) (2018) 075122.
\newblock \href {https://doi.org/10.1103/PhysRevB.98.075122}
  {\path{doi:10.1103/PhysRevB.98.075122}}.

\bibitem{benali_toward_2020}
A.~Benali, K.~Gasperich, K.~D. Jordan, T.~Applencourt, Y.~Luo, M.~C. Bennett,
  J.~T. Krogel, L.~Shulenburger, P.~R.~C. Kent, P.-F. Loos, A.~Scemama,
  M.~Caffarel, Toward a systematic improvement of the fixed-node approximation
  in diffusion {Monte} {Carlo} for solids—{A} case study in diamond, J. Chem.
  Phys. 153~(18) (2020) 184111, publisher: American Institute of Physics.
\newblock \href {https://doi.org/10.1063/5.0021036}
  {\path{doi:10.1063/5.0021036}}.

\bibitem{wagner_transition_2007}
L.~K. Wagner, Transition metal oxides using quantum {Monte} {Carlo}, J. Phys.:
  Condens. Matter 19~(34) (2007) 343201, publisher: IOP Publishing.
\newblock \href {https://doi.org/10.1088/0953-8984/19/34/343201}
  {\path{doi:10.1088/0953-8984/19/34/343201}}.

\bibitem{dubecky_noncovalent_2016}
M.~Dubecký, L.~Mitas, P.~Jurečka, Noncovalent {Interactions} by {Quantum}
  {Monte} {Carlo}, Chem. Rev. 116~(9) (2016) 5188--5215, publisher: American
  Chemical Society.
\newblock \href {https://doi.org/10.1021/acs.chemrev.5b00577}
  {\path{doi:10.1021/acs.chemrev.5b00577}}.

\bibitem{dash_tailoring_2021}
M.~Dash, S.~Moroni, C.~Filippi, A.~Scemama, Tailoring {CIPSI} {Expansions} for
  {QMC} {Calculations} of {Electronic} {Excitations}: {The} {Case} {Study} of
  {Thiophene}, J. Chem. Theory Comput. 17~(6) (2021) 3426--3434, publisher:
  American Chemical Society.
\newblock \href {https://doi.org/10.1021/acs.jctc.1c00212}
  {\path{doi:10.1021/acs.jctc.1c00212}}.

\bibitem{huang_bandgaps_2021}
X.~Huang, H.~Zhang, X.-L. Cheng, Bandgaps in free-standing monolayer {TiO2}:
  {Ab} initio diffusion quantum {Monte} {Carlo} study, International Journal of
  Quantum Chemistry 121~(12) (2021) e26643.
\newblock \href {https://doi.org/10.1002/qua.26643}
  {\path{doi:10.1002/qua.26643}}.

\bibitem{wines_first-principles_2020}
D.~Wines, K.~Saritas, C.~Ataca, A first-principles {Quantum} {Monte} {Carlo}
  study of two-dimensional ({2D}) {GaSe}, J. Chem. Phys. 153~(15) (2020)
  154704, publisher: American Institute of Physics.
\newblock \href {https://doi.org/10.1063/5.0023223}
  {\path{doi:10.1063/5.0023223}}.

\bibitem{malone_systematic_2020}
F.~D. Malone, A.~Benali, M.~A. Morales, M.~Caffarel, P.~R.~C. Kent,
  L.~Shulenburger, Systematic comparison and cross-validation of fixed-node
  diffusion {Monte} {Carlo} and phaseless auxiliary-field quantum {Monte}
  {Carlo} in solids, Phys. Rev. B 102~(16) (2020) 161104, publisher: American
  Physical Society.
\newblock \href {https://doi.org/10.1103/PhysRevB.102.161104}
  {\path{doi:10.1103/PhysRevB.102.161104}}.

\bibitem{wang_binding_2020}
G.~Wang, A.~Annaberdiyev, L.~Mitas, Binding and excitations in {Si$_x$H$_y$}
  molecular systems using quantum {Monte} {Carlo}, J. Chem. Phys. 153~(14)
  (2020) 144303, publisher: American Institute of Physics.
\newblock \href {https://doi.org/10.1063/5.0022814}
  {\path{doi:10.1063/5.0022814}}.

\bibitem{annaberdiyev_cohesion_2021}
A.~Annaberdiyev, G.~Wang, C.~A. Melton, M.~C. Bennett, L.~Mitas, Cohesion and
  excitations of diamond-structure silicon by quantum {Monte} {Carlo}:
  {Benchmarks} and control of systematic biases, Phys. Rev. B 103~(20) (2021)
  205206, publisher: American Physical Society.
\newblock \href {https://doi.org/10.1103/PhysRevB.103.205206}
  {\path{doi:10.1103/PhysRevB.103.205206}}.

\bibitem{melton_spin-orbit_2016}
C.~A. Melton, M.~Zhu, S.~Guo, A.~Ambrosetti, F.~Pederiva, L.~Mitas, Spin-orbit
  interactions in electronic structure quantum {Monte} {Carlo} methods, Phys.
  Rev. A 93~(4) (2016) 042502.
\newblock \href {https://doi.org/10.1103/PhysRevA.93.042502}
  {\path{doi:10.1103/PhysRevA.93.042502}}.

\bibitem{melton_quantum_2016}
C.~A. Melton, M.~C. Bennett, L.~Mitas, Quantum {Monte} {Carlo} with variable
  spins, J. Chem. Phys. 144~(24) (2016) 244113.
\newblock \href {https://doi.org/10.1063/1.4954726}
  {\path{doi:10.1063/1.4954726}}.

\bibitem{melton_fixed-node_2016}
C.~A. Melton, L.~Mitas, Fixed-{Node} and {Fixed}-{Phase} {Approximations} and
  {Their} {Relationship} to {Variable} {Spins} in {Quantum} {Monte} {Carlo},
  in: Recent {Progress} in {Quantum} {Monte} {Carlo}, Vol. 1234 of {ACS}
  {Symposium} {Series}, American Chemical Society, 2016, pp. 1--13.
\newblock \href {https://doi.org/10.1021/bk-2016-1234.ch001}
  {\path{doi:10.1021/bk-2016-1234.ch001}}.

\bibitem{melton_quantum_2017}
C.~A. Melton, L.~Mitas, Quantum {Monte} {Carlo} with variable spins:
  {Fixed}-phase and fixed-node approximations, Phys. Rev. E 96~(4) (2017)
  043305.
\newblock \href {https://doi.org/10.1103/PhysRevE.96.043305}
  {\path{doi:10.1103/PhysRevE.96.043305}}.

\bibitem{hu_many-body_2012}
S.~Hu, K.~Rasch, L.~Mitas, Many-{Body} {Nodal} {Hypersurface} and {Domain}
  {Averages} for {Correlated} {Wave} {Functions}, in: Advances in {Quantum}
  {Monte} {Carlo}, Vol. 1094 of {ACS} {Symposium} {Series}, American Chemical
  Society, 2012, pp. 77--87, section: 7.
\newblock \href {https://doi.org/10.1021/bk-2012-1094.ch007}
  {\path{doi:10.1021/bk-2012-1094.ch007}}.

\bibitem{li_atomic_2011}
X.~Li, J.~Kolorenč, L.~Mitas, Atomic {Fermi} gas in the unitary limit by
  quantum {Monte} {Carlo} methods: {Effects} of the interaction range, Phys.
  Rev. A 84~(2) (2011) 023615, publisher: American Physical Society.
\newblock \href {https://doi.org/10.1103/PhysRevA.84.023615}
  {\path{doi:10.1103/PhysRevA.84.023615}}.

\bibitem{giner_using_2013}
E.~Giner, A.~Scemama, M.~Caffarel, Using perturbatively selected configuration
  interaction in quantum {Monte} {Carlo} calculations, Can. J. Chem. 91~(9)
  (2013) 879--885, publisher: NRC Research Press.
\newblock \href {https://doi.org/10.1139/cjc-2013-0017}
  {\path{doi:10.1139/cjc-2013-0017}}.

\bibitem{lopez_rios_inhomogeneous_2006}
P.~López~Ríos, A.~Ma, N.~D. Drummond, M.~D. Towler, R.~J. Needs,
  Inhomogeneous backflow transformations in quantum {Monte} {Carlo}
  calculations, Phys. Rev. E 74~(6) (2006) 066701, publisher: American Physical
  Society.
\newblock \href {https://doi.org/10.1103/PhysRevE.74.066701}
  {\path{doi:10.1103/PhysRevE.74.066701}}.

\bibitem{bajdich_pfaffian_2006}
M.~Bajdich, L.~Mitas, G.~Drobný, L.~K. Wagner, K.~E. Schmidt, Pfaffian
  {Pairing} {Wave} {Functions} in {Electronic}-{Structure} {Quantum} {Monte}
  {Carlo} {Simulations}, Phys. Rev. Lett. 96~(13) (2006) 130201, publisher:
  American Physical Society.
\newblock \href {https://doi.org/10.1103/PhysRevLett.96.130201}
  {\path{doi:10.1103/PhysRevLett.96.130201}}.

\bibitem{bajdich_pfaffian_2008}
M.~Bajdich, L.~Mitas, L.~K. Wagner, K.~E. Schmidt, Pfaffian pairing and
  backflow wavefunctions for electronic structure quantum {Monte} {Carlo}
  methods, Phys. Rev. B 77~(11) (2008) 115112, publisher: American Physical
  Society.
\newblock \href {https://doi.org/10.1103/PhysRevB.77.115112}
  {\path{doi:10.1103/PhysRevB.77.115112}}.

\bibitem{townsend_starting-point-independent_2020}
J.~P. Townsend, S.~D. Pineda~Flores, R.~C. Clay, T.~R. Mattsson, E.~Neuscamman,
  L.~Zhao, R.~E. Cohen, L.~Shulenburger, Starting-point-independent quantum
  {Monte} {Carlo} calculations of iron oxide, Phys. Rev. B 102~(15) (2020)
  155151, publisher: American Physical Society.
\newblock \href {https://doi.org/10.1103/PhysRevB.102.155151}
  {\path{doi:10.1103/PhysRevB.102.155151}}.

\bibitem{luchow_direct_2007}
A.~Lüchow, R.~Petz, T.~C. Scott, Direct optimization of nodal hypersurfaces in
  approximate wave functions, J. Chem. Phys. 126~(14) (2007) 144110, publisher:
  American Institute of Physics.
\newblock \href {https://doi.org/10.1063/1.2716640}
  {\path{doi:10.1063/1.2716640}}.

\bibitem{toulouse_optimization_2007}
J.~Toulouse, C.~J. Umrigar, Optimization of quantum {Monte} {Carlo} wave
  functions by energy minimization, J. Chem. Phys. 126~(8) (2007) 084102,
  publisher: American Institute of Physics.
\newblock \href {https://doi.org/10.1063/1.2437215}
  {\path{doi:10.1063/1.2437215}}.

\bibitem{hilbert_mathematical_1902}
D.~Hilbert, Mathematical problems, Bull. Amer. Math. Soc. 8~(10) (1902)
  437--479.
\newblock \href {https://doi.org/10.1090/S0002-9904-1902-00923-3}
  {\path{doi:10.1090/S0002-9904-1902-00923-3}}.

\bibitem{umrigar_diffusion_1993}
C.~J. Umrigar, M.~P. Nightingale, K.~J. Runge, A diffusion {Monte} {Carlo}
  algorithm with very small time‐step errors, J. Chem. Phys. 99~(4) (1993)
  2865--2890, publisher: American Institute of Physics.
\newblock \href {https://doi.org/10.1063/1.465195}
  {\path{doi:10.1063/1.465195}}.

\bibitem{sogge_2011}
C.~D. Sogge, S.~Zelditch, Lower bounds on the {Hausdorff} measure of nodal
  sets, Math. Res. Lett. 18 (2011) 25.

\bibitem{stanke_five_2009}
M.~Stanke, J.~Komasa, S.~Bubin, L.~Adamowicz, Five lowest ${^{1}S}$ states of
  the {Be} atom calculated with a finite-nuclear-mass approach and with
  relativistic and {QED} corrections, Phys. Rev. A 80~(2) (2009) 022514,
  publisher: American Physical Society.
\newblock \href {https://doi.org/10.1103/PhysRevA.80.022514}
  {\path{doi:10.1103/PhysRevA.80.022514}}.

\bibitem{dolg_relativistic_2012}
M.~Dolg, X.~Cao, Relativistic {Pseudopotentials}: {Their} {Development} and
  {Scope} of {Applications}, Chem. Rev. 112~(1) (2012) 403--480, publisher:
  American Chemical Society.
\newblock \href {https://doi.org/10.1021/cr2001383}
  {\path{doi:10.1021/cr2001383}}.

\end{thebibliography}





\end{document}